\documentstyle[aps,prl,twocolumn,psfig]{revtex}
\begin{document}
\draft
\title{THEORY OF OPTICAL TWEEZERS}
\author{P. A. Maia Neto and H. M. Nussenzveig}
\address{Instituto de F\'{\i}sica, Universidade Federal do Rio de Janeiro, Caixa\\
Postal 68528,\\
21945-970 Rio de Janeiro, Rio de Janeiro, Brazil}

\date{\today }

\maketitle

\begin{abstract}
We derive a partial-wave (Mie) expansion of the axial force exerted
on a transparent sphere by a laser beam focused through a high numerical
aperture objective. 
The results hold throughout the range of interest for practical
applications. The ray optics
limit is shown to follow from the Mie expansion by size averaging. Numerical
plots show large deviations from ray optics near the focal region and oscillatory
behavior (explained in terms of a simple interferometer picture) of the force as a function of the size parameter. Available experimental data favor the present model over previous ones.
\end{abstract}


Optical tweezers are single-beam laser traps for neutral particles that have
a wide range of applications in physics and biology~\cite{1}. Dielectric
microspheres are trapped and employed as handles in most of the quantitative
applications. The gradient trapping force is applied by bringing the laser
beam to a diffraction limited focal spot through a large numerical aperture
microscope objective.

Typical size parameters $\beta = ka$ ($a =$ microsphere radius, $k =$ laser
wavenumber) range in order of magnitude from values $< 1$ to a few times $%
10^1.$ A theory of the trapping force based on geometrical optics (GO)~\cite
{2} should not work in this range. Other proposals (cf.~\cite{1}), based on
Mie theory, have employed unrealistic near-paraxial models for the
transverse laser beam structure near the focus, incompatible with its large
angular aperture.

We take for the incident beam {\it before}
 the objective, propagating along the
positive $z$ axis, the usual Gaussian ${\rm (TEM)_{00}}$ transverse laser
mode profile, with beam waist $w_0$ at the input aperture, where $k w_0\gg
1. $ We employ the Richards and Wolf~\cite{3} representation for the
corresponding strongly focused beam {\it beyond} the objective, with a large
opening angle $\theta_0$ (no paraxial assumption), taking due account of the
Abbe sine condition. This should be a more realistic representation.

The microsphere, with real refractive index $n_2$ (we neglect absorption),
is immersed in a homogeneous medium with refractive index $n_1.$ We consider
here the simplest situation, in which the sphere center is aligned with the
laser beam axis, so that we evaluate the {\it axial} trapping force. With
origin at the sphere center, we denote by ${\bf r} = -q \hat {{\bf z}}$ the
focal point position. The fraction $A$ of total beam power that enters the
lens aperture is
\begin{equation}
A=1-\exp (-2\gamma ^2\sin ^2\theta _0).  \label{A}
\end{equation}
where $\gamma$ is the ratio of the objective focal length to the beam waist $%
w_0.$

By axial symmetry, the trapping force in this situation is independent of
input beam polarization: we take circular polarization. The electric field
of the strongly focused beam (we omit the time factor $\exp (-i\omega t)$)
has the Debye-type~\cite{3} integral representation 
\[
{\bf E}_0({\bf r})=E_0\int_0^{2\pi }d\phi \int_0^{\theta _0}d\theta \sin
\theta \sqrt{\cos \theta }\exp \left( -\gamma ^2\sin ^2\theta \right) 
\]
\begin{equation}
\times \exp \left[ i{\bf k}\cdot ({\bf r}+q\hat{{\bf z}})\right] 
\mbox{\boldmath $\hat{\epsilon}$}(\theta,\phi),  \label{incident}
\end{equation}
where $k=|{\bf k}(\theta ,\phi )|=n_1\omega /c,$ $ \mbox{\boldmath $\hat{\epsilon}$}(\theta,\phi)={\hat{{\bf x}}}^{\prime }+i{\hat{{\bf y}}}^{\prime },$ and
the unit vectors ${\hat{{\bf x}}}^{\prime }$ and ${\hat{{\bf y}}}^{\prime }$
are obtained from $\hat{{\bf x}}$ and $\hat{{\bf y}},$ respectively, by
rotation with Euler angles $\alpha =\phi ,\beta =\theta ,\gamma =-\phi .$
The factor $\sqrt{\cos\theta}$ arises from the Abbe sine condition. 

For each plane wave $\exp(i{\bf k}\cdot{\bf r})$ in the superposition~(\ref
{incident}), the corresponding scattered field is given by the well-known
Mie partial-wave series~\cite{4}, in terms of the Mie coefficients $a_l,
b_l, $ that are functions of the size parameter $\beta$ and the relative
refractive index $n=n_2/n_1.$ By substitution into~(\ref{incident}), we
obtain the total scattered field ${\bf E}_s({\bf r}).$

The trapping force is found by replacing the total field ${\bf E}={\bf E}_0
+ {\bf E}_s$ (likewise for ${\bf B}$) into the Maxwell stress tensor and
integrating over the surface of the sphere. The resulting axial force $F$ is
proportional to the focused laser beam power $P,$ 
\begin{equation}
F = (n_1/c) P Q,  \label{defi}
\end{equation}
where $Q$ is the (dimensionless) axial trapping efficiency~\cite{1}.

We denote by $Q_e$ the contribution from terms in $E_0E_s$ and $B_0B_s$
(that also give rise to the extinction efficiency) and by $Q_s$ the
remaining terms, so that $Q=Q_e+Q_s.$ We find 
\begin{equation}
Q_e={\frac{4\gamma ^2}A}{\rm Re}\sum_{l=1}^\infty
(2l+1)(a_l+b_l)G_lG_l{}^{^{\prime }}{}^{*},  \label{main1}
\end{equation}
\[
Q_{s}= {\frac{-8\gamma^2}{A}} {\rm Re} \Biggl[\sum_{l=1}^{\infty} {\frac{%
l(l+2)}{l+1}} (a_{l} a_{l+1}^* + b_{l} b_{l+1}^*)G_{l} G_{l+1}^*
\]
\begin{equation}
+ {\frac{(2l+1)}{l(l+1)}} a_{l}b_{l}^*G_{l}G_{l}^* \Biggr], \label{main2}
\end{equation}
where $G_{l}$ and $G_{l}{}^{^{\prime}}$ are multipole coefficients for the
focused beam, 
\[
G_{l}=\int_0^{\theta_0} d\theta\sin\theta \sqrt{\cos\theta} 
\exp(-\gamma^2
\sin^2\theta) \exp(i \delta\cos\theta)
\]
\begin{equation}
\times\, d_{1,1}^{l}(\theta),  \label{G}
\end{equation}
\begin{equation}
G_l{}^{^{\prime}} = -i \partial G_l/\partial \delta,  \label{G'}
\end{equation}
with $\delta = k q.$ In~(\ref{G}), $d_{1,1}^{l}(\theta)$ are the matrix
elements of finite rotations~\cite{5}, that can be expressed as 
\begin{equation}
d_{1,1}^{l}(\theta) = \left[p_l(\cos\theta) + t_l(\cos\theta)\right]/(2l+1)
\end{equation}
in terms of the Mie angular functions~\cite{6} $p_l$ and $t_l.$ The results~(%
\ref{main1}) and (\ref{main2}), apart from converging beam effects, have the
same structure as the radiation pressure efficiency~\cite{6}, with which
they are closely related, as will be seen below.

In the Rayleigh limit, $\beta\ll 1,$ $Q$ is dominated by the electric dipole
Mie term $a_1,$ and the trapping force~(\ref{defi}) becomes $F=(\alpha/2) 
{\bf \nabla E^2},$ where $\alpha$ is the static polarizability of the sphere~%
\cite{1}. In the opposite limit $\beta\gg 1,$ the connection with
geometrical optics is established by applying to (\ref{main1}) and (\ref
{main2}) the following steps~\cite{6}~\cite{7}. (i)~In (6), substitute $p_l$
and $t_l$ by their (non-uniform) asymptotic expansions for large $l,$ and
approximate $G_l$ and $G_l^{^{\prime}}$ by the method of stationary phase~%
\cite{8}. (ii) Compute the average $<Q>$ over a size parameter range
associated with a quasiperiod of the Mie coefficients. The result is 
\[
<Q>_{{\rm GO}}= {\frac{4\gamma^2}{A}} \int_0^{\theta_0} d\theta
\sin\theta\cos\theta\exp(-2\gamma^2\sin^2\theta)
\]
\[
\times
\Biggl\{\cos\theta+ {\frac{1}{2}}\sum_{j=1}^2 r_j\cos(2\theta_1-\theta)
\]
\begin{equation}
 -{\frac{1}{2}}{\rm Re} \sum_{j=1}^2 (1-r_j)^2 {\frac{e^{i[2(\theta_1-%
\theta_2)-\theta]}}{1+r_j e^{-2i\theta_2}}}\Biggr\}. 
\label{geo}
\end{equation}
In (\ref{geo}), $\theta_1=\arcsin(q\sin\theta/a),$ $\theta_2=\arcsin(\sin%
\theta_1/n)$ are the angles of incidence and refraction (defined so as to be
negative if $q<0$) at the sphere surface associated with a component in the
direction $\theta$ of the focused beam (\ref{incident}). The corresponding
Fresnel reflectivity for polarization $j$ ($\|,$ $\bot$) is $r_j.$ Eq.~(\ref{geo}) may also
be derived in the framework of GO. Thus, the expression within curly
brackets agrees with the GO result for the force exerted by each component
ray as first obtained in~\cite{9}. The remaining pre-factors in (\ref{geo}),
not accounted for previously, represent the intensity 
distribution of the focused beam
as implied by the sine condition and the transverse profile of the laser
beam at the input aperture of the objective.

In Fig.~1, $Q$ is plotted as a function of $q/a,$ the center offset from the
focus in units of the sphere radius~\cite{num}. The numerical values chosen
correspond to the experiment of Ref.~\cite{10}: $n_1 =1.33, n_2 = 1.57, A =
0.85, \theta_0 = 78^{\rm o},$ 
which by~(\ref{A}) yield $\gamma^2 = 0.99. $ The
dotted curve represents the GO result~(\ref{geo}). The other curves
represent the exact Mie results~(\ref{main1}) and (\ref{main2}) for two
different $\beta$ values, corresponding to microsphere radii employed in~%
\cite{10}: $1.42 \mu {\rm m}$ (dashed) and $2.16 \mu {\rm m}$ (solid), respectively $%
\beta = 18.8$ and $\beta = 28.4.$ The qualitative behavior of the GO curve
has been explained~\cite{2} in terms of competition between radiation
pressure (scattering force) and gradient force. However, the GO result for
the maximum backward trapping efficiency
$Q_m$ is smaller (by a factor of the order of $2$%
)~\cite{f1} than the values obtained in Ref.~\cite{2}. This is 
in line with the discrepancy between experimental and theoretical values noted
in Ref.~\cite{block}.

The Mie theory provides values for $Q_m$ 
($0.088$ for $\beta=18.8$ and $0.086$
for $\beta=28.4$~\cite{f2}) below the GO result $Q_m=0.095.$ The position at
which the backward force is maximum lies
beyond the corresponding GO value $%
q/a=1.01$. The stiffness decreases as this point is approached from the
focus, contradicting the GO prediction and in agreement with an experiment
reported in Ref.~\cite{block}.

GO is also a poor approximation near the geometrical focus, as expected. In
fact, Fig.~1 shows that the exact values deviate substantially
 from GO near $q=0.$ The stable equilibrium 
position shows large positive as well as negative 
offsets from GO (further discussed below), and
the linear Hooke's law
range around the equilibrium position is narrower than
predicted by GO. Because of the axial focusing effect~\cite{6}, the
(nonuniform) asymptotic approximations to the Mie angular functions employed
in the derivation of~(\ref{geo}) break down at $q = 0,$ although~(\ref{geo})
is continuous at this point, yielding, with $\theta_1=\theta_2=0,$ 
\begin{equation}
<Q>_{{\rm GO}}(q=0)={\frac{4 r}{1+r}} <\cos\theta>,  \label{q01}
\end{equation}
where $r$ is the Fresnel reflectivity for normal incidence and $<\cos\theta>$
denotes an average over the intensity distribution of the focused beam~(\ref
{incident}). Since the incident rays are either backscattered or undeviated
in this approximation, (\ref{q01}) represents pure radiation pressure in GO.

The region around $q = 0$  deserves special treatment, in view of its
relevance to the evaluation of trap axial 
stiffness. For $\beta \gg 1,$ the above
discussion and the localization principle imply that the main contributions
to (\ref{main1}) and (\ref{main2}) should arise from partial waves with $%
l\ll \beta , $ so that we apply Hankel's asymptotic expansions to the
spherical Bessel functions in the Mie coefficients. The results are
independent of $l,$ and the summations over multipole coefficients can then
be carried out, resulting in 
\begin{equation}
Q(q=0)={\frac{8r\sin ^2\Delta /2}{1+r^2-2r\cos \Delta }}<\cos \theta >,
\label{q02}
\end{equation}
where $\Delta =4n_2\omega a/c.$  This expression corresponds to the
radiation pressure 
efficiency
(twice the reflectivity) of an infinite set of
parallel-plate interferometers 
(width $2a,$ refractive index $n_2,$ so that $%
\Delta $ is the round-trip phase), each one oriented at an angle $\theta $
with respect to the axis, 
traversed at normal incidence by the 
corresponding beam angular component. 
The GO result~(\ref{q01}) follows from~(\ref{q02})
by taking an incoherent average. Since $n-1$ is small, we have $r\ll 1,$ so
that the interferometer reflectivity is nearly sinusoidal.

In Fig.~2, for the same parameters as in Fig.~1, we plot $Q$ at $q = 0$ as a
function of $\beta.$ The Mie curve (full line) displays the expected
near-sinusoidal oscillation as $\beta$ increases, approaching the
interferometer behavior~(\ref{q02}) (shown in dotted line). The GO value~(%
\ref{q01}) (dashed line) is approached in the average sense. The two points
corresponding to the $\beta$ values employed in Fig.~1 are shown by circles.
Since the radiation pressure at $\beta=28.4$ is above the GO value, the Mie
value for the equilibrium position $q_{{\rm eq}}$ is larger than the GO
result, in agreement with Fig.~1 (the opposite applies at $\beta=18.8$). The
values for $q_{{\rm eq}}$ are found by numerically solving the equation $%
Q(q_{{\rm eq}})=0.$ In the limit $\beta\gg 1,$ they are vanishingly small at 
$\beta$ values that are minima of $Q(q=0).$ For $\beta \stackrel{>}{%
\scriptstyle\sim }5,$ $q_{{\rm eq}}/a$ as a 
function of $\beta$ oscillates in
phase with the oscillations of $Q(q = 0),$ around the GO value $(q_{{\rm eq}%
}/a)_{{\rm GO}}=0.217,$ and with amplitude of the order of $0.17.$

The trap axial stiffness is given by 
\begin{equation}
\kappa =-{\frac{n_1P}c} \left({\frac{\partial Q}{\partial q}}\right)_{q=q_{%
{\rm eq}}}.  \label{kappa1}
\end{equation}
Within GO, $\kappa$ decreases as $1/\beta.$ This follows from scaling: $Q_{%
{\rm GO}}$ depends on $q$ only through $q/a.$ Hence, $\partial Q_{{\rm GO}%
}/\partial q=Q_{{\rm GO}}^{\prime }(q/a)/a,$ yielding 
\begin{equation}
\kappa _{{\rm GO}}=-{\frac{n_1P}c}Q_{{\rm GO}}^{\prime } 
\!\!\left({\frac {q_{{\rm eq}%
}}a}\right){\frac k\beta }.  \label{kappa2}
\end{equation}
Again for the parameters of Ref.~\cite{10} (power $P = 3 {\rm mW}$), we plot in
Fig.~3 the Mie values of $\kappa$  (solid line)~\cite{footnote}, the
GO result $\kappa_{\rm GO}
= (18/\beta)({\rm pN}/{\rm \mu m})$ calculated from Eq.~(%
\ref{kappa2}) (dotted line) and the experimental data points from Ref.~\cite
{10}, with the respective error bars. We also show (dot-dashed line) the
values predicted by the electrostatic model recently suggested by Tlusty et
al.~\cite{14}. As could be expected, their approach may be applied only in
the low-frequency (Rayleigh) limit, where it may be replaced by the simpler
electric 
dipole approximation (neglecting the variation of the 
field over the sphere
volume) already discussed above 
in connection with~(\ref{main1}). In order to
test the sensitivity of the results to the focused beam parameters, we also
plot the Mie values for $\kappa$ corresponding to a larger waist: $\gamma^2
= 0.3$ (dashed line).

For $\beta \stackrel{>}{\scriptstyle\sim }10,$ the Mie values for $\kappa $
oscillate around the GO curve with period $\Delta \beta =\pi /(2n),$ like
the force $Q(q=0)$ [cf.~(\ref{q02})] and the equilibrium position. This
corresponds to a frequency interval $\Delta\nu = c/(4n_2 a),$ which is in
the THz range for spheres with radii of a few microns. As shown in the inset
in Fig.~3, where we plot $(<\kappa >-\kappa _{{\rm GO}})/\kappa _{{\rm GO}}$
as function of $\beta ,$ the average of the Mie values, $<\kappa >,$ stands
above the GO curve, but the relative difference decreases to zero as $\beta $
increases beyond $\beta \approx 55.$ For large sphere diameters, $\kappa $
may become very small over short $\beta $ intervals. This may be of interest
for applications to scanning force microscopy~\cite{10}.

In conclusion, by deriving an analytic Mie expansion for the axial trapping
efficiency, based on a more realistic model of optical tweezers, we are
able to cover the range of interest for most applications. Furthermore,
the connection with the correct GO limit (taking into account 
the sine condition)
has been derived from the Mie
expansion by size averaging. The behavior near the focus has  been
obtained and interpreted in terms of an interferometer model, which also
accounts for equilibrium position and trap axial 
stiffness oscillations. These
oscillations should be accessible to experiments 
by scanning the laser beam
frequency. Most 
experimental data points lie above the GO values, closer to the
wave optics predictions computed from  the Mie expansion employing only
the experimentally given parameters.

We thank W. Wiscombe for useful suggestions and programs for quadrature
integration and for Mie scattering calculations, and CNPq for partial
support. One of us (P. A. M. N.) acknowledges support by Programa de
N\'ucleos de Excel\^encia (PRONEX), grant 4.1.96.08880.00-7035-1.

\begin{figure}
\psfig{file=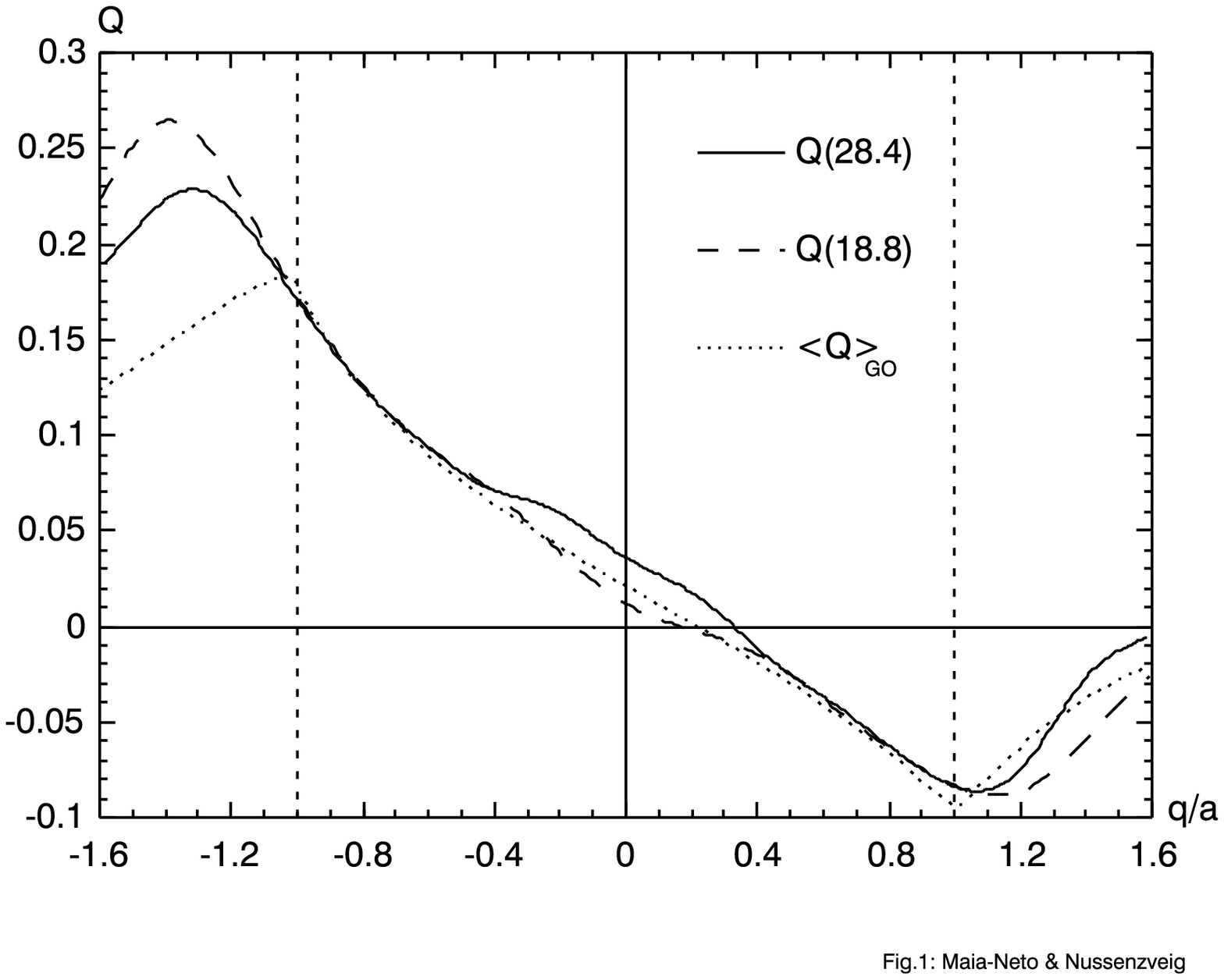,height=6.4cm,width=8.5cm}
\caption{Normalized axial 
force versus position (in units of the sphere 
radius). The dotted 
line is computed from ray optics theory, whereas the solid and dashed
lines are calculated from the wave--optics theory with size parameters $%
\beta= 28.4$ and $\beta=18.8,$ respectively. The vertical dashed lines 
mark the microsphere boundaries.}
\end{figure}

\begin{figure}
\psfig{file=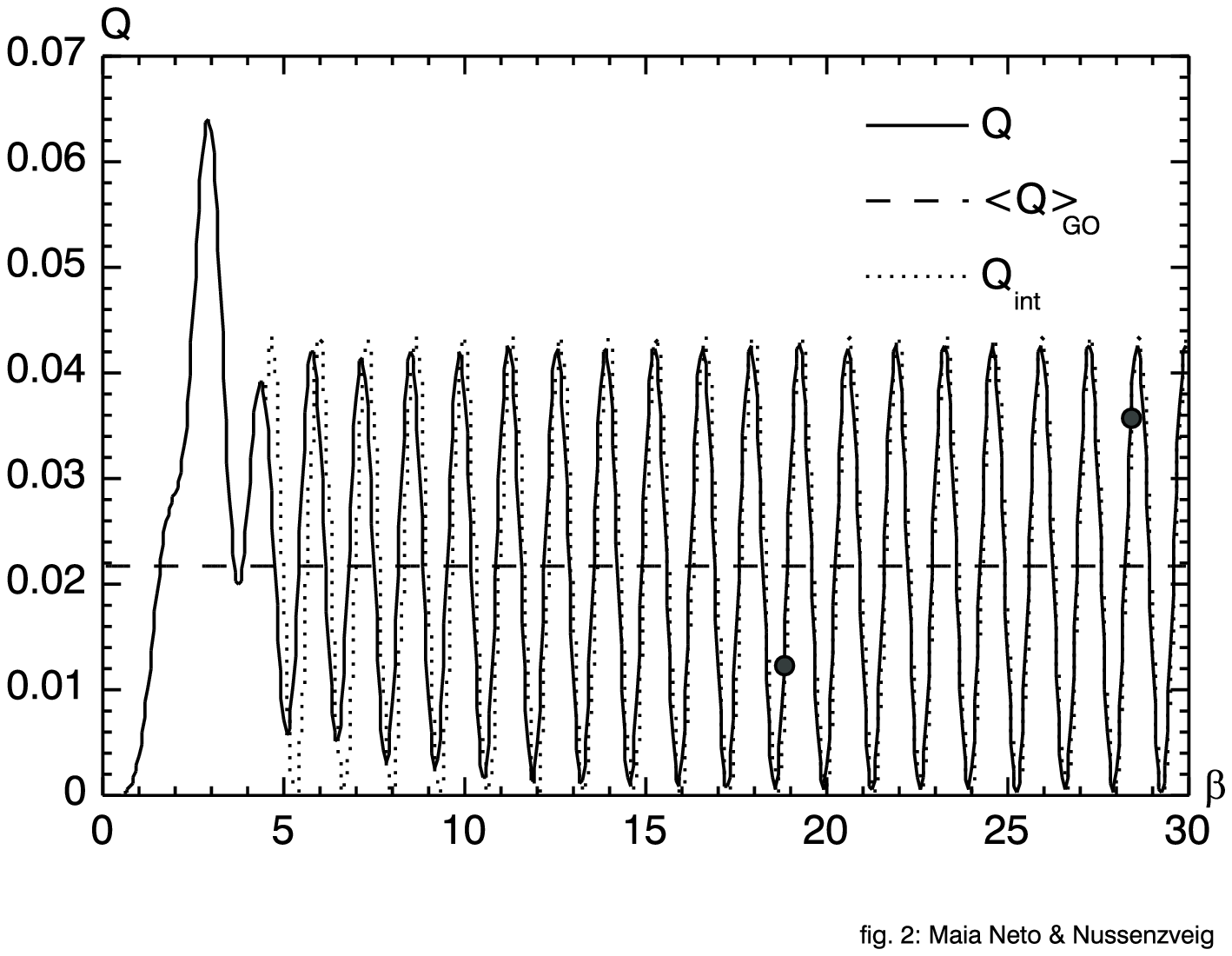,height=6.4cm,width=8.5cm}
\caption{Normalized 
force at the geometrical focal point versus size parameter $\beta:$ exact (full line); interferometer model (dotted line) and geometrical optics
(horizontal dashed line). 
The black  circles 
indicate the two values of $\beta $ used in Fig.~1.}
\end{figure}

\begin{figure}
\psfig{file=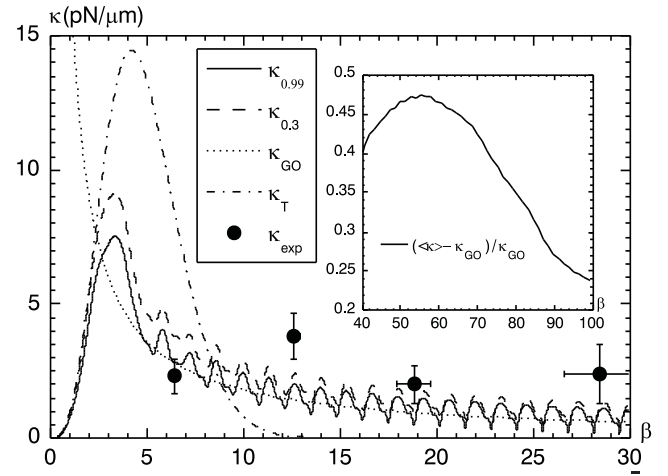,height=6.4cm,width=8.5cm}
\caption{Axial 
stiffness $\kappa$ of the optical tweezer as a function of $\beta.$
Solid, dotted and dot-dashed lines correspond to the (exact) wave--optics
theory, geometrical optics, and 
electrostatic theory~[16], 
respectively, and 
for a focal length to waist squared ratio  $\gamma^2=0.99.$ 
Also shown are
the experimental data points of Ref.~[11], with
corresponding error bars, and 
the exact values for $\gamma^2=0.3$ (dashed line).
 In the inset, we plot
the relative discrepancy between the average of the 
exact values  
and the geometrical optics results (for $\gamma^2=0.99$).}
\end{figure}


\begin{references}
\bibitem{8}  The stationary-phase point ($q=0$ is excluded) is at $\bar{%
\theta}=\arcsin [(l+1/2)/k|q|],$ as expected by the localization principle~%
\cite{6}.
\bibitem{num}  The numerical integrations in Eqs.~(\ref{G}) and (\ref{geo})
were performed with the help of a Kronrod-Patterson adaptative Gaussian-type
quadrature method.
\bibitem{f1}  The gradient force is overestimated in~\cite{2} as a
consequence of neglecting the sine condition and the corresponding factor $%
\cos \theta $ in Eq.~(\ref{geo}), which diminishes the contribution of rays
at large angles. By the same reason, the stable equilibrium positions as
predicted by (\ref{geo}) are further from the focus than the values obtained
in~\cite{2}.
\bibitem{footnote} The Mie values for $\kappa$ are 
obtained by deriving from (\ref{main1}) and (\ref{main2}) the 
partial-wave expansion for $\partial_q Q,$ and then replacing 
the results for $q_{\rm eq}$ into (\ref{kappa1}).
\end{references}
\end{document}